\documentclass[12pt]{article}

\usepackage[ruled,vlined]{algorithm2e}
\usepackage{amsfonts}
\usepackage{amsmath}
\usepackage{amssymb}
\usepackage{amsthm}
\usepackage{dsfont}
\usepackage{bm}
\usepackage{array}
\usepackage{authblk}
\usepackage{bm}
\usepackage{booktabs}
\usepackage{caption}
\usepackage{color}
\usepackage{graphicx}
\usepackage{multirow}
\usepackage[authoryear,semicolon]{natbib}
\usepackage{physics}
\usepackage{siunitx}
\usepackage{standalone}
\usepackage{subfigure}
\usepackage{tikz}
\usepackage[normalem]{ulem}
\usepackage{url}
\usepackage[colorlinks,linkcolor=blue,citecolor=blue,urlcolor=blue]{hyperref}

\newcolumntype{x}[1]{>{\centering\arraybackslash\hspace{0pt}}p{#1}}
\newcommand{\Var}{\text{Var}}

\newcommand{\omegavec}{\boldsymbol{\omega}}
\newcommand{\phivec}{\boldsymbol{\phi}}
\newcommand{\betavec}{\boldsymbol{\beta}}
\newcommand{\epsilonvec}{\boldsymbol{\epsilon}}
\newcommand{\etavec}{\boldsymbol{\eta}}

\newcommand{\thetavec}{\boldsymbol{\theta}}

\newcommand{\psivec}{\boldsymbol{\psi}}
\newcommand{\zerovec}{\boldsymbol{0}}

\newcommand{\Pivec}{\boldsymbol{\Pi}}
\newcommand{\Phivec}{\boldsymbol{\Phi}}
\newcommand{\Sigmavec}{\boldsymbol{\Sigma}}

\newcommand{\fvec}{\mathbf{f}}

\newcommand{\hvec}{\mathbf{h}}

\newcommand{\svec}{\mathbf{s}}

\newcommand{\wvec}{\mathbf{w}}
\newcommand{\xvec}{\mathbf{x}}

\newcommand{\zvec}{\mathbf{z}}

\newcommand{\Gau}{\text{Gau}}

\newcommand{\Avec}{\mathbf{A}}

\newcommand{\Dvec}{\mathbf{D}}
\newcommand{\Ivec}{\mathbf{I}}

\newcommand{\Xvec}{\mathbf{X}}
\newcommand{\Yvec}{\mathbf{Y}}
\newcommand{\Zvec}{\mathbf{Z}}

\theoremstyle{definition}

\usepackage{stackengine}
\stackMath
\newcommand\tsup[2][2]{%
 \def\useanchorwidth{T}%
  \ifnum#1>1%
    \stackon[-.5pt]{\tsup[\numexpr#1-1\relax]{#2}}{\scriptscriptstyle\sim}%
  \else%
    \stackon[.5pt]{#2}{\scriptscriptstyle\sim}%
  \fi%
}

\usepackage{geometry}
\begin{document}

\def\spacingset#1{\renewcommand{\baselinestretch}%
{#1}\small\normalsize} \spacingset{1.5}


  \title{\Large{\textbf{deepspat: An \textsf{R} package for modeling nonstationary spatial and spatio-temporal Gaussian and extremes data through deep deformations}}}
  
  \author[1]{Quan Vu}
  \author[2]{Xuanjie Shao}
  \author[2]{Raphaël Huser}
  \author[3]{Andrew Zammit-Mangion}

  \affil[1]{\small{Research School of Finance, Actuarial Studies and Statistics, Australian National University, Australia}}
  \affil[2]{\small{Statistics Program, CEMSE Division, King Abdullah University of Science and Technology, Saudi Arabia}}
  \affil[3]{\small{School of Mathematics and Applied Statistics, University of Wollongong, Australia}}

  \date{}
  \maketitle

\begin{abstract}

Nonstationarity in spatial and spatio-temporal processes is ubiquitous in environmental datasets, but is not often addressed in practice, due to a scarcity of statistical software packages that implement nonstationary models. In this article, we introduce the \textsf{R} software package \textbf{deepspat}, which allows for modeling, fitting and prediction with nonstationary spatial and spatio-temporal models applied to Gaussian and extremes data. The nonstationary models in our package are constructed using a deep multi-layered deformation of the original spatial or spatio-temporal domain, and are straightforward to implement. Model parameters are estimated using gradient-based optimization of customized loss functions with \textbf{tensorflow}, which implements automatic differentiation. The functionalities of the package are illustrated through simulation studies and an application to Nepal temperature data.

\end{abstract}

\noindent
{\it Keywords}: Deep learning, Gaussian process, Max-stable process, Pareto process, Warping. 

\spacingset{1.5} 


\newpage

\section{Introduction}

In recent years, many different statistical methods have been introduced to tackle the challenge of modeling and fitting large spatial and spatio-temporal data. Along with these, several software packages have been developed to fit complex spatial and spatio-temporal models and predict an underlying process at unobserved locations.
The traditional approach to modeling geostatistical (that is, point-referenced) data involves Gaussian processes, which do not naturally scale to modeling big data. 
A comprehensive review of approaches to deal with this issue can be found in \cite{heaton2019case}. Some examples of these methods include fixed rank kriging, implemented in the \textsf{R} package \textbf{FRK} \citep{sainsbury2024modeling}, nearest neighbor Gaussian process, with the package \textbf{spNNGP} \citep{finley2022spnngp}, and the stochastic partial differential equation approach, often combined with the integrated nested Laplace approximation technique implemented in the \textsf{R} package \textbf{INLA} \citep{lindgren2015bayesian}.
Another issue arising from large spatial and spatio-temporal data is that of capturing nonstationarity. Stationarity refers to the assumption that the marginal and dependence characteristics are the same everywhere throughout the spatial/spatio-temporal domain (that is, they are invariant with respect to spatial/temporal locations); however, stationarity is often violated when the process is observed over a large or topographically complex domain. 
There are a number of approaches to address nonstationarity. For example, nonstationary covariance functions can be constructed through kernel convolutions, as implemented in the package \textbf{convoSPAT} \citep{risser2017local}. 
Another parametric approach allows the parameters of the covariance function to be covariate-dependent, as implemented in the package \textbf{cocons} \citep{blasi2024class}. 
Yet another class of approaches involves spatial deformation, where the spatial domain is warped to induce approximate nonstationarity \citep{sampson1992nonparametric}. Alternatively, the spatial domain can be expanded in dimension to flexibly capture nonstationary dependence \citep{bornn2012modeling}. These deformation/expansion approaches are implemented in the \textsf{R} package \textbf{deform} \citep{youngman2023deform}.
However, \textbf{deform} has some limitations: i) it only supports Gaussian processes, and requires multiple replicates as it uses an empirical covariance matrix when estimating the deformation; 
ii) computation is limited to a small number of spatial locations; and iii) bijectivity is enforced via grid penalties and is tuning-sensitive. 
Recently, nonparametric approaches have also been used for covariance estimation, which naturally capture nonstationarity \citep[e.g.,][]{kidd2022bayesian}. 
In spatial data analysis, sometimes one also needs to model spatial extremes \citep{davison2012statistical, davison2015statistics, huser2022advances}, to assess risk and understand the behavior of rare but impactful events such as extreme temperatures, precipitation, and wind speeds. Classical spatial extremes modeling often involves the use of max-stable processes \citep{ribatet2013spatial} or generalized Pareto processes \citep{de2018high} to capture the dependence structure among extreme values across space. 
Several approaches have been proposed to address nonstationarity in spatial extremal dependence. These include those of \cite{huser2016non} and \cite{zhong2022modeling}, who incorporate covariates into models of the underlying extremal dependence structure, \cite{castro2020local}, who employ local likelihood methods to capture spatial variation, and \cite{shao2025flexible}, who use a (penalized) regularization approach based on max-stable processes. Another strategy involves regionalization via clustering techniques tailored to extremes \citep{bernard2013clustering, maume2024regionalization}. Spatial deformation frameworks based on splines have also been investigated in this setting; 
see \cite{smith1996estimating} and \cite{richards2021spatial}. However, these spline-based methods often suffer from space folding issues and computational inefficiencies. Bayesian hierarchical models have also been developed for spatially structured inference driven by latent processes \citep{cooley2007bayesian}, but apart from a few exceptions \citep[e.g.,][]{reich2012hierarchical,bopp2021hierarchical}, they usually do not properly account for data-level dependence, let alone capture nonstationarity in the dependence structure (but see \citealp{zhang2023efficient}). 

A number of \textsf{R} packages have been developed to facilitate inference for spatial extremes; see, e.g., \cite{belzile2023modeler} for a recent comprehensive overview. For instance, the package \textbf{SpatialExtremes} \citep{ribatet2013spatial} provides tools for fitting max-stable processes via pairwise likelihood methods, while \textbf{evgam} \citep{youngman2022evgam} allows for the modeling of univariate extremes using generalized additive models with flexible spatial components. 
For incorporating spatially-structured covariate effects in univariate extremes, packages such as \textbf{evd} \citep{stephenson2002evd}, \textbf{ismev} \citep{heffernan2016ismev}, and \textbf{extRemes} \citep{gilleland2016extremes} can also be used.
More recently, the \textbf{mvPot} package \citep{de2021mvpot} has been introduced to enable spatial peaks-over-threshold modeling, offering inference tools particularly suited to estimating the spatial dependence structure in extreme events through $r$-Pareto processes. However, all of these approaches and packages are limited in their ability to flexibly or realistically capture spatial nonstationarity and/or in their applicability to high-dimensional data, and there is currently no available software to model nonstationarity in the dependence structure of spatial extreme processes.

In this paper, we introduce the \textsf{R} package \textbf{deepspat} for modeling spatial and spatio-temporal data using nonstationary process models. The underlying methodology was developed originally for Gaussian processes by \cite{zammit2022deep}, and subsequently adapted to the context of multivariate spatial \citep{vu2021modeling} or spatio-temporal \citep{vu2022constructing} Gaussian processes, and spatial extremes \citep{shao2025modeling}. The underlying approach is to infer flexible deformations (also referred to as warpings) of the spatial domain as a way to construct nonstationary processes \citep{sampson1992nonparametric}. The deformation function maps the original spatial/spatio-temporal domain to a new warped domain, such that the transformed process is stationary on this new domain. The deformation function is assumed to admit a multi-layered form to capture the complexity of the nonstationary behavior, and is constructed from bijective warping units to avoid space folding. Our package relies on minimizing customized loss functions and utilizing the functionality of \textbf{tensorflow} \citep{TensorflowR}, which implements automatic differentiation to compute gradients of the loss function with respect to model parameters, and which facilitates gradient-based optimization with models that have a large number of parameters.
The package allows for fitting and predicting with Gaussian data, including univariate spatial and spatio-temporal data and multivariate spatial data, through Gaussian processes. The package also implements the nearest neighbor Gaussian process \citep{datta2016hierarchical}, which has a sparse precision matrix instead of a dense precision matrix, and fixed rank kriging \citep{cressie2008fixed}, which uses spatial basis functions. These modeling approaches reduce the computational burden of model fitting when the number of observed data points is large. 
In addition to Gaussian models, \textbf{deepspat} can also be used for modeling nonstationary spatial extremes---namely, 
max-stable and $r$-Pareto processes of Brown--Resnick type, for spatial block maxima and high threshold exceedances, respectively.

The remainder of the article is organized as follows. In Section~\ref{sec:method}, we provide a statistical overview of the models implemented in the \textbf{deepspat} package. This includes the technical details of the deformation approach used to construct the nonstationary models, and those for model fitting and prediction with Gaussian and extreme-value processes. In Section~\ref{sec:package}, we give a brief summary of package functionalities, along with a simple simulation example. In Section~\ref{sec:application}, we present an application of \textbf{deepspat} to a real-world temperature dataset for a mountainous region over Nepal, and demonstrate our package for modeling the bulk and extremes of temperature using Gaussian and max-stable processes, respectively. Section~\ref{sec:conclusion} concludes with an outlook on possible future software developments.


\section{Methodology}\label{sec:method}

\subsection{Spatial covariance functions and variograms}
A central consideration of spatial and spatio-temporal modeling is how to account for the data's dependence structure. For ease of exposition consider the spatial-only case, and denote a spatial process model by $Y(\svec), \svec \in \mathcal{S} \subset \mathbb{R}^k, k \in \{2,3\}$, over a spatial domain $\mathcal{S}$, where $\Var(Y(\cdot)) < \infty$. 
Classical spatial modeling usually proceeds by specifying the mean function, $\mathbb{E}\{Y(\svec)\}, \svec \in \mathcal{S}$, and the covariance function of the process on $\mathcal{S}$, defined as
\begin{equation*}
    C\left(\svec_1, \svec_2\right) = \operatorname{Cov}\left\{Y(\svec_1), Y(\svec_2)\right\} = \mathbb{E}\left[\left\{Y\left(\svec_1\right)-\mathbb{E}\left(Y(\svec_1)\right)\right\}\left\{Y\left(\svec_2\right)-\mathbb{E}\left(Y(\svec_2)\right)\right\}\right],
\end{equation*}
where $\svec_1, \svec_2 \in \mathcal{S}$. To simplify spatial modeling, it is common to assume that $Y(\cdot)$ is stationary and isotropic.

The covariance function is \emph{stationary} when it only depends on the displacement between the locations, $\hvec = \svec_1 - \svec_2$, that is, when one can write
\begin{equation}\label{eq:cov}
    C(\svec_1, \svec_2) \equiv C^{o}(\hvec),
\end{equation}
for some suitable function $C^{o}:\mathbb{R}^k\to\mathbb{R}$. 
The assumption of stationarity implies that the second-order behavior of the process is shift-invariant on $\mathcal{S}$. More generally, the dependence structure may instead be characterized through the semivariogram 
\begin{equation*}
    \gamma(\svec_1, \svec_2)=\frac{1}{2}\operatorname{Var}\{Y(\svec_1)-Y(\svec_2)\}. 
\end{equation*}
As in \eqref{eq:cov}, the semivariogram is called stationary if one can write
\begin{equation}
    \gamma(\svec_1, \svec_2) \equiv \gamma^{o}(\hvec),
\end{equation}
for $\hvec = \svec_1 - \svec_2$ and some suitable function $\gamma^{o}:\mathbb{R}^k\to\mathbb{R}^+$. 
In this case, the process of interest is said to be intrinsically stationary. Note that if the covariance function is stationary, then the semivariogram is also stationary, but the converse is not necessarily true.

Another ubiquitous modeling assumption renders the process rotation-invariant. A stationary and \emph{isotropic} covariance function is one that only depends on the distance between sites, that is, one for which 
\begin{equation}
    C\left(\svec_1, \svec_2\right) \equiv C^{o}\left(h\right),
\end{equation}
where $h = \|\hvec\| = \|\svec_1 - \svec_2 \|$ for all $\svec_1, \svec_2 \in \mathcal{S}$, for some norm $\|\cdot\|$ over $\mathcal{S}$ (most often the Euclidean norm) and some suitable function ${C^{o}:\mathbb{R}^+\to\mathbb{R}}$. Similarly, the assumption of isotropy may be imposed when working with the semivariogram, by letting $\gamma \left(\svec_1, \svec_2\right) \equiv \gamma^{o}\left(h\right)$, 
where ${\gamma^{o}:\mathbb{R}^+\to\mathbb{R}^+}$.

In this package, we employ the spatial deformation approach to model nonstationary dependence structures. To do so, we rely on two components: a deformation (or warping) function ${\fvec: \mathcal{S} \to \mathcal{W} \subset \mathbb{R}^k}$ of the spatial domain, and a stationary and isotropic covariance/variogram structure on the warped domain $\mathcal{W}$.
In other words, we model nonstationarity by constructing (and estimating) flexible warpings $\fvec(\cdot)$, such that 
$C\left(\svec_1, \svec_2\right) = C^{o}\left( \norm{\fvec(\svec_1) - \fvec(\svec_2)} \right)$ or $\gamma\left(\svec_1, \svec_2\right) = \gamma^{o}\left( \norm{\fvec(\svec_1) - \fvec(\svec_2)} \right)$. In the spatio-temporal case, the warping function warps space and time separately, and we model a stationary (but possibly non-separable) covariance function on the warped spatio-temporal domain.
We next describe how flexible warpings are constructed in \textbf{deepspat}.

\subsection{Deformations}\label{sec:deformations}

\begin{figure}[t!]
\begin{center}
\includegraphics[width=0.99\linewidth]{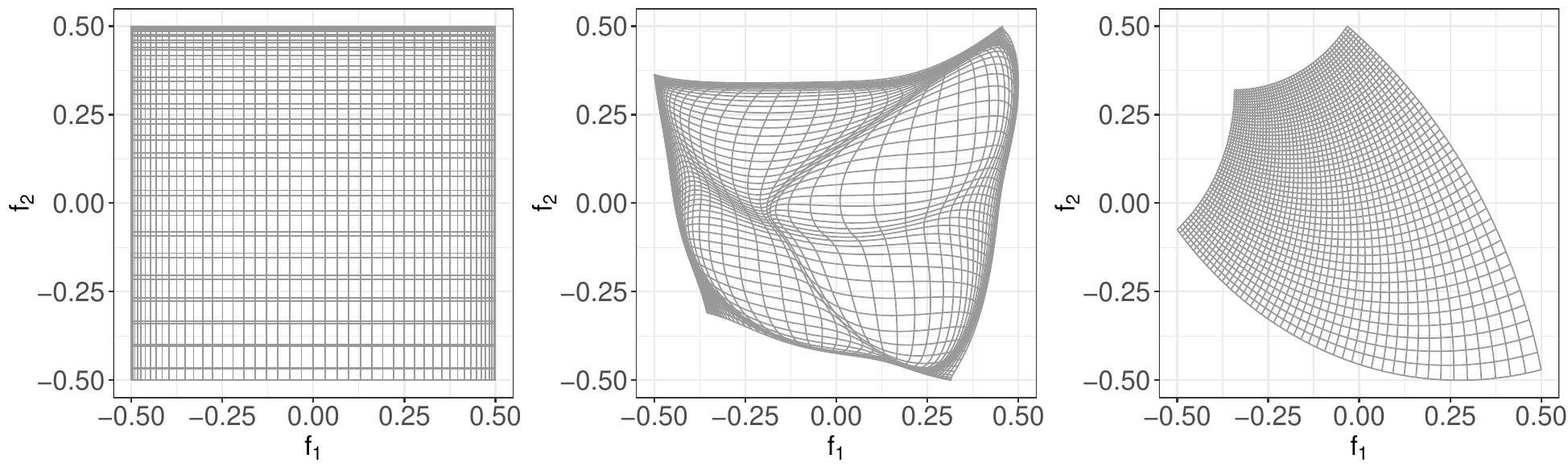}
\end{center}
\caption{Action of an axial warping unit (left), radial basis function unit (middle), and M\"{o}bius transformation unit (right), when applied to the original space represented by a regular grid.}
\label{fig:warp}
\end{figure}

Consider again the spatial-only case. We model the deformation (or warping) function as an injective function that maps the spatial coordinates from a geographical domain $\mathcal{S}$ to a new warped domain $\mathcal{W}$. To make the deformation function highly flexible, we model it as a composition of $L$ simple injective warping units (or layers), $\fvec_1(\cdot), \dots, \fvec_L(\cdot)$, resulting in a function of the form \citep{zammit2022deep},
\begin{equation}\label{eq:deformation}
\fvec(\cdot) = \fvec_L \circ \dots \circ \fvec_1(\cdot).
\end{equation}
In \textbf{deepspat}, the types of warping units that users can implement are axial warping units, radial basis function units, and M\"{o}bius transformation units. These are visualized in Figure~\ref{fig:warp}. 
An axial warping (AW) unit is used to stretch or contract one spatial coordinate at a time using a linear combination of one linear basis function and $(r-1)$ sigmoid basis functions, given by
\begin{equation}\label{eq:AWU}
f(s) = \sum_{i = 1}^{r} \omega_i \varphi_i(s),
\end{equation}
where $s$ is one of the spatial coordinates in $\svec$, and where 
\begin{equation}\label{eq:AWU_2}
\varphi_1(s) = s \quad \textrm{and} \quad \varphi_i(s) = \frac{1}{1 + \exp{-b_1(s-b_{2i})}}, \quad i = 2, \dots, r.
\end{equation}
The parameters that need to be estimated in an AW unit are the basis function weights, $\omega_i, {i = 1, \dots, r}$. Users can specify $b_1 \in \mathbb{R}^{+}$, which controls the steepness of the sigmoid basis functions, and $r \in \mathbb{N}$, the number of basis functions. The other hyperparameters, $b_{2i}, i = 2, \dots, r$, represent the sigmoid basis function centers, and are automatically chosen in the software to be equally spaced (given the value of $r$).

A radial basis function (RBF) unit is a two-dimensional warping function given by
\begin{equation}\label{eq:RBF}
\fvec(\svec) = \svec + \omega \exp{-b \norm{\svec - \mathbf{c}}^2} (\svec - \mathbf{c}).
\end{equation}
An RBF stretches or contracts space in the vicinity of the center of the basis function.
All parameters in an RBF are fixed, except for the weight $\omega$, which needs to be estimated. In the package, RBFs are available in units at different resolutions; the first resolution of RBFs is $3 \times 3$ on a square domain, while the second resolution is $9 \times 9$, and so on.
Users can specify the resolution of the unit, which then automatically determines the number of basis functions used and the locations $\mathbf{c}$ of their centers, and also the hyperparameter $b$, such that the unit can smoothly warp the domain.

A M\"{o}bius transformation (MT) unit, which is also a two-dimensional unit, is given by
\begin{equation}
\fvec(\svec) = \begin{pmatrix} \Re(\varphi(\svec)) \\ \Im(\varphi(\svec)) \end{pmatrix}, 
\end{equation}
where
\begin{equation*}
\varphi(\svec) = \frac{\omega_1 z(\svec) + \omega_2}{\omega_3 z(\svec) + \omega_4}; \quad z(\svec) = s_1 + i s_2 ; \quad \omega_1, \omega_2, \omega_3, \omega_4 \in \mathbb{C}.
\end{equation*}
An MT unit is used to induce large-scale deformation.
The parameters that need to be estimated are the real and imaginary parts of the parameters $\omega_1, \omega_2, \omega_3, \omega_4$ (eight parameters in total). Note that a MT is also a linear fractional transformation, and in \textbf{deepspat} the associated function is given by \texttt{LFT()}.

For the spatio-temporal case, space and time are warped separately. Space is warped as described above, while time, being one-dimensional, is warped using an AW unit.  

\subsection{Nonstationary models for Gaussian data}\label{sec:gaussian_models}

The package \textbf{deepspat} allows the user to model two types of data: spatial/spatio-temporal Gaussian data and spatial extremes. Here we give details of the former case.

Suppose that we have $n$ real-valued observations $Z_1, \dots, Z_n$ of a spatial process at locations $\svec_1,\dots,\svec_n \in \mathcal{S}$; we model these as
\begin{equation}\label{eq:Gaussian_model}
    Z_i = \xvec'(\svec_i) \betavec + Y(\svec_i) + \epsilon_i, \quad i = 1, \dots, n,
\end{equation}
where $\xvec(\cdot): \mathcal{S} \rightarrow \mathbb{R}^q, q \ge 1$, is a vector containing covariates, $Y(\cdot)$ is a zero-mean nonstationary Gaussian process, and $\epsilon_i$ is an independent Gaussian error term.
The Gaussian process $Y(\cdot)$ has a nonstationary covariance structure, modeled through the warping function \eqref{eq:deformation}, and a stationary isotropic covariance structure $C^{o}(\cdot)$ on the warped domain $\mathcal{W}$. Specifically, we define the covariance function on the original domain $\mathcal{S}$ as
\begin{equation}\label{eq:nonstat_structure_cov}
C(\svec_1, \svec_2) = C^{o}(\norm{\fvec(\svec_1) - \fvec(\svec_2)}; \psivec), \quad \svec_1, \svec_2 \in \mathcal{S},
\end{equation}
where $\psivec$ are covariance function parameters that need to be estimated. Note from \eqref{eq:Gaussian_model} that the likelihood function of these parameters for any given $\Zvec=(Z_1,\ldots,Z_n)'$ has a Gaussian form.
In \textbf{deepspat}, the parameter vector $\thetavec = (\psivec',\omegavec')'$, which comprises the covariance function parameters $\psivec$ and warping weights and parameters $\omegavec$, 
are estimated by maximizing the log restricted likelihood, given by
\begin{equation}\label{eq:reml}
\mathcal{L}(\thetavec; \Zvec) = -\frac{n - q}{2} \log(2 \pi) + \frac{1}{2} \log \abs{\Xvec' \Xvec} + \frac{1}{2} \log \abs{\Sigmavec_{\Zvec}^{-1}} - \frac{1}{2} \log \abs{\Xvec' \Sigmavec_{\Zvec}^{-1} \Xvec} - \frac{1}{2} \Zvec' \Pivec \Zvec,
\end{equation}
where
\( \Pivec = \Sigmavec_{\Zvec}^{-1} - \Sigmavec_{\Zvec}^{-1} \Xvec (\Xvec' \Sigmavec_{\Zvec}^{-1} \Xvec)^{-1} \Xvec' \Sigmavec_{\Zvec}^{-1} \), $\Sigmavec_{\Zvec} = \Sigmavec_{\Yvec} + \Sigmavec_{\epsilonvec}$, and $\Sigmavec_{\Yvec}, \Sigmavec_{\epsilonvec}$ are the covariance matrices of $\Yvec = (Y(\svec_1), \dots, Y(\svec_n))'$ and $\epsilonvec = (\epsilon_1, \dots, \epsilon_n)'$, respectively. The covariance matrix $\Sigmavec_{\Yvec}$ is constructed using the nonstationary covariance function defined in equation \eqref{eq:nonstat_structure_cov}. After estimating $\thetavec$, the vector of regression coefficients $\betavec$ is estimated using generalized least squares. 

When the number of observations $n$ is large, evaluating the log-likelihood function in \eqref{eq:reml} becomes computationally infeasible, mainly due to the computation related to inverting the covariance matrix $\Sigmavec_{\Zvec}$. There are a number of approaches to modify the model \eqref{eq:Gaussian_model} to enable fast and efficient computation, two of which are implemented in \textbf{deepspat}. The first employs a nearest-neighbor Gaussian process model, while the second employs a basis-function model.

The nearest-neighbor Gaussian process model provides a sparse approximation to Gaussian processes \citep{finley2019efficient}. 
In this model, the value of the process at each location is modeled to be conditionally dependent given a subset of observations. Specifically, using $p(\cdot)$ to denote generic (conditional) distribution functions, the distribution of $\Zvec$ is approximated by
\begin{equation}
p(\Zvec) \approx \prod_{i=1}^{n} p(Z_i \mid \Zvec_{N(\svec_i)}),
\end{equation}
where $N(\svec_i)$ is the set of indices of the nearest neighbors (out of the locations from $\svec_1$ to $\svec_{i-1}$) of $\svec_i$. As a result, the precision matrix, that is, the inverse of the covariance matrix $\Sigmavec_{\Zvec}$, can be approximated as $\Sigmavec_{\Zvec}^{-1} \approx (\Ivec - \Avec)' \Dvec^{-1} (\Ivec - \Avec),$ where $\Avec$ is a sparse matrix, and $\Dvec$ is a diagonal matrix. For more details, see \cite{finley2019efficient}.

The second, basis-function modeling approach, also known as fixed rank kriging, is a widely used approach in spatial statistics \citep[e.g.,][]{cressie2008fixed, sainsbury2024modeling, zammit2021frk}. Here, we model the underlying process as a basis-function model constructed as
\begin{equation}
    Y(\svec) = \phivec'(\fvec(\svec)) \etavec,
\end{equation}
where $\phivec(\cdot)$ is a vector of pre-specified spatial basis functions (where the number of basis functions is much smaller than the number of observations), and $\etavec$ is a vector of random weights. We model $\etavec \sim \Gau(\zerovec, \Sigmavec_{\etavec})$, where $\Sigmavec_{\etavec}$ is the covariance matrix of the weights, which depends on the distance between the centroids of the basis functions. These centroids are chosen on the warped space $\mathcal{W}$.
Then, the covariance matrix of $\Yvec = (Y(\svec_1), \ldots, Y(\svec_n))'$ is $(\phivec(\svec_i)' \Sigmavec_{\etavec} \phivec(\svec_j): i,j = 1,\dots,n) = \Phivec \Sigmavec_{\etavec} \Phivec'$, where $\Phivec = (\phivec(\svec_i): i =1,\dots,n)'$, and the precision matrix of $\Zvec$ is therefore
\begin{equation}
\Sigmavec_{\Zvec}^{-1} =  (\Sigmavec_{\Yvec} + \Sigmavec_{\epsilonvec})^{-1} = \Sigmavec_{\epsilonvec}^{-1} - \Sigmavec_{\epsilonvec}^{-1} \Phivec (\Sigmavec_{\etavec}^{-1} + \Phivec' \Sigmavec_{\epsilonvec}^{-1} \Phivec)^{-1} \Phivec' \Sigmavec_{\epsilonvec}^{-1},
\end{equation}
which does not involve inverting large dense matrices since $\Sigmavec_{\etavec}$ is low-dimensional and $\Sigmavec_{\epsilonvec}$ is diagonal.

In \textbf{deepspat}, one can have two variants of \eqref{eq:Gaussian_model} when modeling Gaussian data. The first variant is a spatio-temporal Gaussian process model $Y(\svec, t)$, which allows the user to model nonstationarity in both space and time \citep{vu2022constructing}. For example, the nonstationary spatio-temporal covariance function of $Y(\cdot, \cdot)$ can be defined to be separable and modeled as
$$
C(\svec_1, \svec_2; t_1, t_2) = C_S^{o}(\norm{\fvec_{\svec}(\svec_1) - \fvec_{\svec}(\svec_2)}) C_T^{o}(\abs{f_t(t_1) - f_t(t_2)}),
$$
where $\fvec_{\svec}(\cdot)$ is a spatial warping function, $f_t(\cdot)$ is a temporal warping function, and $C_S^{o}:\mathbb{R}^+\to\mathbb{R}$ and $C_T^{o}:\mathbb{R}^+\to\mathbb{R}$ are spatial and temporal stationary isotropic covariance functions, respectively.

The second variant is a multivariate spatial Gaussian process $\Yvec(\svec) = (Y_1(\svec), \dots, Y_p(\svec))'$, where $p$ is the number of spatial processes \citep[the current implementation in the package allows $p$ = 2 or 3;][]{vu2021modeling}. The cross-covariance function for the $p_1$-th and $p_2$-th processes is modeled as
$$
C_{p_1 p_2}(\svec_1, \svec_2) = \text{Cov}\{Y_{p_1}(\svec_1), Y_{p_2}(\svec_2) \} = C^{o}_{p_1 p_2}(\norm{ \fvec_{p_1}(\svec_1) - \fvec_{p_2}(\svec_2)}),
$$
where $\fvec_{p}(\cdot)$ represents the spatial warping for the $p^{\text{th}}$ process and $C^{o}_{p_1 p_2}:\mathbb{R}^+\to\mathbb{R}$ is an appropriate stationary isotropic cross-covariance function for the $p_1$-th and $p_2$-th processes on the warped domain.

\subsection{Nonstationary models for extremes data}\label{sec:Ext}


Suppose now that $Y(\svec), \svec\in\mathcal{S}$, is a stochastic process with marginal distribution function $F_{\svec}(\cdot)$, assumed to be continuous. In extremes, we often model margins and extremal dependence structure separately from the bulk. To focus on extremal dependence, we consider without loss of generality, the standardized process $X(\svec), \svec \in \mathcal{S}$, defined as $X(\svec) = [1-F_{\svec}\{Y(\svec)\}]^{-1}$, which has standard Pareto margins, i.e., $\textrm{Pr}\{X(\svec)\leq x\}=1-1/x$, $x>1$. Assume that functional extremes of $X$, in a sense made precise below, converge to some nondegenerate limiting process $Z$, whose extremal dependence structure is characterized by a nonstationary semivariogram $\gamma(\cdot, \cdot)$. In \textbf{deepspat}, nonstationary extremal processes are then constructed through the warping function \eqref{eq:deformation}, and a stationary isotropic semivariogram structure $\gamma^{o}(\cdot)$ on the warped domain $\mathcal{W}$, given by
\begin{equation}\label{eq:nonstat_structure_vario}
    \gamma(\svec_1, \svec_2) = \gamma^{o}(\|\fvec(\svec_1) - \fvec(\svec_2)\|; \psivec), \quad \svec_1, \svec_2 \in \mathcal{S},
\end{equation}
where $\psivec$ are parameters that need to be estimated. There are multiple ways to define spatial extremes. In \textbf{deepspat} we model nonstationary extremal dependence using two types of extremal processes: max-stable processes \citep[see, e.g.,][]{brown1977extreme, smith1990max, schlather2002models, opitz2013extremal} for spatial block maxima, and $r$-Pareto processes \citep[see, e.g.,][]{dombry2015functional, de2018high} for spatial peaks-over-threshold. 

In this section, $Y(\cdot)$ represents the true random process that is directly observed, $X(\cdot)$ is the random process on a standardized unit Pareto scale, and $Z(\cdot)$ denotes a limiting extreme-value process on which the analysis is carried out, while corresponding lowercase notation is used for realized values. This is standard notation in the spatial extremes literature; but note that it redefines the notation used in the Gaussian process Section~\ref{sec:gaussian_models}, where $Y(\cdot)$ is a latent process (with noise removed), $\xvec(\cdot)$ denotes a vector of covariates, and $Z_1,\dots, Z_n$ are noisy measurements..


\noindent \underline{\textbf{Nonstationary max-stable processes}}

Let $\{X_i(\svec): i=1,\dots,B\}$ be independent and identically distributed (i.i.d.) copies of $X(\svec)$, $\svec \in \mathcal{S}$ within a time block with size $B$. One can show that the renormalized pointwise maximum process $B^{-1}\max\{X_1(\svec), \ldots, X_B(\svec)\}$ converges to a (simple) max-stable process $Z(\svec)$ with unit Fr\'echet margins, i.e., $\textrm{Pr}\{Z(\svec)\leq z\}=\exp(-1/z)$, $z>0$. This process may be represented as \citep{de1984spectral}
\begin{equation}
    Z(\svec) = \sup_{j\geq 1} W_j(\svec)/\Gamma_j
    \label{maxstab},
\end{equation}
where $\{\Gamma_j\}_{j\geq 1}$ are points of a Poisson process on $(0, \infty)$ with unit rate intensity and $\{W_j(\svec)\}$, ${\svec\in \mathcal{S}}$, are i.i.d.\ copies of a nonnegative stochastic process satisfying the condition $\mathbb{E}\{W(\svec)\} = 1$. The $n$-dimensional joint distribution of $Z(\svec)$ at the sites $\svec_1, \ldots, \svec_n \in \mathcal{S}$ is
\begin{equation}
\operatorname{Pr}\left\{Z_{1} \leq z_{1}, \ldots, Z_{n} \leq z_{n}\right\}=\exp \left\{-V\left(z_{1}, \ldots, z_{n}\right)\right\},
\label{maxsta_cdf}
\end{equation}
where $Z_i = Z\left(\svec_{i}\right), i=1,\ldots,n$, and $V(\cdot)$ is the exponent function. Explicit formulas for the exponent function exist only for specific choices of $W(\cdot)$. One example is the Brown--Resnick model proposed by \cite{brown1977extreme} and \cite{kabluchko2009}. It is defined by specifying $W(\svec) = \exp\{U(\svec) - \sigma^2(\svec)/2\}$ in (\ref{maxstab}), where $U(\svec), \svec\in\mathcal{S}$, is a centered Gaussian process with semivariogram $\gamma(\cdot, \cdot): \mathcal{S} \times \mathcal{S} \rightarrow [0,\infty)$ and variance function $\sigma^2(\svec)$. The stationary isotropic power semivariogram $\gamma^o(h) = (h/\varphi)^{\kappa}$, with smoothness parameter $\kappa\in(0,2]$ and range parameter $\varphi>0$, is a common choice, which we also implement in \textbf{deepspat}. With this model, the bivariate exponent function $V(\cdot)$ has the form
\begin{equation} \label{eq:V}
V\left(z_{i}, z_{j}\right)=\dfrac{1}{z_{i}} \Phi\left\{\dfrac{a}{2}-\dfrac{1}{a} \log \left(\dfrac{z_{i}}{z_{j}}\right)\right\}+\dfrac{1}{z_{j}} \Phi\left\{\dfrac{a}{2}-\dfrac{1}{a} \log \left(\dfrac{z_{j}}{z_{i}}\right)\right\},
\end{equation}
where $a=\sqrt{2\gamma(\svec_i, \svec_j)}$, and $\Phi(\cdot)$ is the standard normal distribution function. There also exists an expression for the $n$-dimensional exponent function \citep[see][]{huser2013composite} but inference becomes prohibitively expensive for moderate-to-large $n$ when this is used \citep{padoan2010likelihood, castruccio2016high, huser2019full}.
To measure extremal dependence between $Z_i$ and $Z_j$, where $i,j = 1,\ldots,n$, we use the extremal coefficient $\vartheta_{ij} = 2\Phi[\sqrt{\gamma(\svec_i, \svec_j) / 2}]\in [1,2]$ \citep{schlather2003dependence}, 
where the strength and decay of extremal dependence within the process is controlled by the nonstationary semivariogram $\gamma(\cdot, \cdot)$.
The case $1 \leq \vartheta_{ij} < 2$ corresponds to asymptotic dependence with decreasing strength of dependence as $\vartheta_{ij}$ increases, and $\vartheta_{ij} = 2$ corresponds to perfect independence between $Z_i$ and $Z_j$. 



\noindent \underline{\textbf{Nonstationary $r$-Pareto processes}}

\cite{dombry2015functional} studied functional threshold exceedances of $X:=\{X(\svec), \svec\in\mathcal{S}\}$, defined as extreme events such that $r(X)\geq u$ for some homogeneous risk functional $r(\cdot)$ and a large threshold $u>1$. Under regularity conditions, $X/u$ given that $r(X)\geq u$ converges, as $u\rightarrow \infty$, to a limit process $Z(\cdot)$ known as an $r$-Pareto process, constructed as:
\begin{equation} \label{eq:Pareto}
    Z(\svec) = PQ(\svec),\quad \svec \in \mathcal{S},
\end{equation}
where $P$ is a unit Pareto random variable, and $Q(\cdot)$ denotes an independent nonnegative stochastic process satisfying $r(Q)=1$. The \textbf{deepspat} package implements the following risk functionals: $r_{\text{max}}(X):=\max_{\svec\in\mathcal{S}'}X(\svec)$, $r_{\text{sum}}(X):=\sum_{\svec\in\mathcal{S}'}X(\svec)$, and $r_{\text{site}}(X):=X(\svec_0)$, where $\mathcal{S}'\subset\mathcal{S}$ denotes a finite set of locations and $\svec_0\in\mathcal{S}$.
A particular class of $r$-Pareto processes that we consider in the package are $r$-Pareto processes associated with Brown--Resnick max-stable processes \citep[see, e.g.,][]{de2018high, dombry2024pareto,zhong2025spatial,shao2025modeling}, where $Q$ in \eqref{eq:Pareto} has a distribution characterized by $\mathbb{P}(Q(\cdot)\in\mathcal{B})=\mathbb{E}[r(W)\mathds{1}\{W(\cdot)/r(W)\in\mathcal{B}\}]/\mathbb{E}[r(W)]$, $W(\cdot)$ is the log-Gaussian process with a stationary isotropic semivariogram $\gamma(\cdot, \cdot)$ defined below~\eqref{maxstab}, and $\mathcal{B}$ is a Borel set of a suitable functions space.

To measure the strength of extremal dependence between two observations from the process observed at two sites $\svec_i, \svec_j\in \mathcal{S}$, we can use the conditional exceedance probability (CEP), defined as
\begin{equation} \label{eq:cep}
\chi_{ij}(u,u')=\mathbb{P}\left[X\left(\svec_j\right) \geqslant u' \mid\left\{X\left(\svec_i\right) \geqslant u'\right\} \cap\left\{r\left(X / u\right) \geqslant 1\right\}\right] 
\end{equation}
for some high thresholds $u, u' > 1$. We can show that for Brown--Resnick processes, the CEP converges to $\chi_{ij} = \lim_{u,u'\rightarrow \infty}\chi_{ij}(u,u')=2\left[1-\Phi\left\{\sqrt{\gamma(\svec_i, \svec_j)/2}\right\}\right]$ when $u'/u\to\infty$.

\noindent \underline{\textbf{Inference for extremal processes}}

Inference for both max-stable and $r$-Pareto processes using the full likelihood function is infeasible with large data sets. We hence use alternative, more computationally efficient objective functions in order to estimate the warping and semivariogram parameters.

\paragraph{Weighted least squares} 
A computationally-efficient way to estimate parameters in the spatial extremes is through weighted least squares. Here, we minimize the discrepancy between the model-based pairwise extremal dependence, indexed by
$\thetavec=(\psivec',\omegavec')'$ and its empirical counterpart. Specifically, we compare extremal coefficients $\vartheta_{ij}(\thetavec)$ (in the max-stable process approach) or conditional extremal probabilities $\chi_{ij}(u,u';\thetavec)$ (in the r-Pareto process approach) to the empirical $\hat{\vartheta}_{ij}$ or $\hat{\chi}_{ij}(u,u')$, respectively, for $\{\svec_i,\svec_j\}\subset\mathcal{S}$. Weights $w_{ij} \geq 0$ (sometimes selected such that $w_{ij}\in\{0,1\}$ for computational efficiency) are employed to downweight weakly dependent pairs and emphasize strongly dependent ones. In \textbf{deepspat}, the weights are specified as $w_{ij}=1/\hat{\vartheta}_{ij}$. In the case of extremal coefficients, the weighted least squares objective is:
\begin{equation}
\ell_{\text{WLS}}(\thetavec) = \sum_{1 \leq i<j \leq n} w_{ij} \left\{ \vartheta_{ij}(\thetavec) - \hat{\vartheta}_{ij} \right\}^2,
\label{eq:WLS_loss}
\end{equation}
where $n$ is the number of spatial locations, and a similar loss function can be defined in terms of conditional exceedance probabilities. Under mild regularity conditions (including parameter identifiability) and under a fixed-warping specification in which $\vartheta_{ij}(\thetavec)$ reduces to $\vartheta_{ij}(\psivec)$, the weighted least squares estimator $\hat{\psivec}_{\mathrm{WLS}}$, which minimizes $\ell_{\text{WLS}}(\psivec)$, is consistent and asymptotically normal as the number of temporal replicates $T \to \infty$ \citep[see the supplementary material of][for details]{shao2025modeling}.
For conditional extremal probabilities $\chi_{ij}(\psivec)$, a similar asymptotic result holds.

\paragraph{Pairwise composite likelihood for parameter estimation with max-stable processes:}
Full likelihood inference for both max-stable processes and $r$-Pareto processes observed at a large number of sites is computationally infeasible. However, one can instead maximize a pairwise composite log‐likelihood built from all bivariate densities. 
Suppose $\mathbf z_t=(z_{1,t},\ldots,z_{n,t})^\top$, $t=1,\ldots,T$, are i.i.d. realizations of the $n$-variate \emph{block-maxima} process obtained from nonoverlapping blocks of size $B$ at sites $\{\svec_i\}_{i=1}^n$, where $z_{i,t}$ denotes the $t$-th block maximum at site $\svec_i$.
Then, one can estimate the parameters $\thetavec$ by minimizing the loss
\begin{equation}
\ell_{\text{PCL}}(\thetavec; \zvec_1, \ldots, \zvec_T) = -\frac{1}{T}\sum_{t=1}^{T} \sum_{1\leq i<j\leq n} w_{ij} \,\ell_{(Z_i, Z_j)}(\thetavec; z_{i,t}, z_{j,t}),
\label{eq:pairwiseLike}
\end{equation}
where $\ell_{(Z_i, Z_j)}(\thetavec; \cdot, \cdot) = \log f_{(Z_i, Z_j)}(\cdot, \cdot; \thetavec)$ is the bivariate log-likelihood for a variable pair $(Z_i, Z_j)$. Similar to the weighted least squares approach, the weights $w_{ij}\geq0$ (often chosen such that $w_{ij}\in\{0,1\}$) are used to downweight the contribution of certain pairs (or completely eliminate them from the pairwise likelihood); partial exclusion of subsets can be used to increase computational and statistical efficiency. In \textbf{deepspat}, the pairwise likelihood weights are specified as $w_{ij}=B_{ij}$, where $B_{ij}, 1\leq i<j\leq n$, are independent Bernoulli random variables with parameter $0<b\leq 1$ controlling the expected proportion of retained pairs.
For max-stable processes with unit Fr\'echet margins, one has $f_{(Z_i, Z_j)}(z_{i,t}, z_{j,t}; \thetavec) = \left\{ V_i(z_{i,t}, z_{j,t})V_j(z_{i,t}, z_{j,t}) - V_{ij}(z_{i,t}, z_{j,t})\right\} \exp\{- V(z_{i,t}, z_{j,t})\}$, where $V_i(z_i,z_j)=\partial V(z_i,z_j)/\partial z_i$, $V_{ij}(z_i,z_j)=\partial^2 V(z_i,z_j)/(\partial z_i\partial z_j)$, and where $z_{i,t}$ denotes the $t$-th block maximum recorded at $\svec_i$ for $i=1,\ldots,n$. Under mild regularity conditions and fixed-warping assumptions in which $\ell_{\text{PCL}}(\thetavec)$ reduces to $\ell_{\text{PCL}}(\psivec)$, the distribution of $\hat{\psivec}_{\text{PCL}}$, the quantity which minimizes $\ell_{\text{PCL}}(\psivec)$, is asymptotically normal for large $T$ \citep{padoan2010likelihood}.


In addition, randomized pairwise (composite) likelihood \citep{mazo2024randomized} is also considered in the \textbf{deepspat} package, where one instead minimizes the loss function
\begin{equation}
\ell_{\mathrm{RPL}}(\thetavec; \zvec_1, \ldots, \zvec_T)\equiv -\frac{1}{T} \sum_{t=1}^T \sum_{1\leq i<j\leq n} B_{t,(i,j)} \ell_{(Z_i, Z_j)}(\thetavec; z_{i,t}, z_{j,t}),
\label{eq:randPairwiseLike}
\end{equation}
where $B_{t,ij}, t=1,\ldots,T, 1\leq i<j\leq n$, are independent Bernoulli random variables with parameter $0<b_T\leq 1$ (dependent on $T$), which act as random binary weights. The distribution of the corresponding estimator of variogram parameters, denoted as $\hat{\psivec}_{\text{RPL}}$, minimizing \eqref{eq:randPairwiseLike} under the fixed-warping assumption, is again asymptotically normal with a similar asymptotic variance that may abbreviate the computation of the costly matrix $\bm{\mathcal{I}}(\cdot)$ as $b_T \to 0$ \citep[see][for details]{mazo2024randomized}.

\paragraph{Gradient score matching for parameter estimation with $r$-Pareto processes:} \cite{de2018high} studied a computationally efficient inference method for $r$-Pareto processes of Brown--Resnick type that utilizes gradient score matching \citep{hyvarinen2005estimation}, since the density of the $r$-Pareto processes cannot be computed easily. Specifically, \cite{de2018high} proposed minimizing a divergence defined by the expected squared distance between the model score function and the data score function, given by 
\begin{equation*}\label{eq:GSM_theo}
\Delta\left(\thetavec\right)=\int_{\mathcal{A}_r}\left\|\nabla_{\zvec} \log \lambda(\zvec; \thetavec) \otimes \wvec(\zvec)-\nabla_{\zvec} \log \lambda(\zvec; \thetavec_0) \otimes \wvec(\zvec)\right\|_2^2 \lambda(\zvec; \thetavec_0) \mathrm{d} \zvec,
\end{equation*}
where $\otimes$ denotes the componentwise product, $\mathcal{A}_r:=\left\{\zvec \in \mathbb{R}_{+}^n: r(\zvec) \geq 1\right\}$ denotes the exceedance region defined by the risk functional $r(\cdot)$, $\thetavec_0$ denotes the true model parameters (in our context, both semivariogram and warping parameters), $\wvec: \mathcal{A}_r \rightarrow \mathbb{R}_{+}^n$ is a positive weight function, and $\lambda(\cdot)$ represents the model's underlying intensity function \citep[see][for its formula corresponding to the Brown--Resnick process]{engelke2015estimation, de2018high}. Here, $\wvec(\zvec) = \{w_i(\zvec)\}_{i=1}^{n}$ is designed to penalize low exceedances when $r(\zvec)$ is close to $u$ with low weights, with the $i$-th component set to $w_i(\zvec) = z_i[1-\exp\{1-r(\zvec)\}]$ by default; some other options are also provided in the \textbf{deepspat} package \citep[see][for details]{de2018high, shao2025modeling}. The corresponding loss function, with the sample evaluation of $\Delta(\thetavec)$ over the warped space, denoted by $\delta(\zvec;\thetavec)$, is given by 
\begin{equation}\label{eq:GSM_loss}
\ell_{\text{GSM}}(\thetavec)=\sum^T_{t=1}\mathds{1}\left\{r\left(\frac{\xvec_t}{u}\right) \geq 1\right\} \delta\left(\frac{\xvec_t}{u}; \thetavec\right).
\end{equation}
Under certain regularity conditions (including parameter identifiability and assumptions about the smoothness of $\wvec(\zvec)$), the estimator $\hat{\psivec}_{\text{GSM}}$ that minimizes \eqref{eq:GSM_loss} under fixed-warping assumption, is asymptotically normal \citep[see][for details]{hyvarinen2005estimation}.

\section{Package overview}\label{sec:package}



\subsection{Tensorflow}

The \textbf{deepspat} package is based on the package \textbf{tensorflow} \citep{TensorflowR}, which is a widely used machine learning package. Inside the \textbf{deepspat} package, a log-likelihood, pairwise log-likelihood, or score-based objective function for each of the models described in Section~\ref{sec:method} is implemented, which is dependent on the model parameters. We leverage automatic differentiation in the \textbf{tensorflow} library to evaluate the gradients of the likelihood function with respect to all model parameters. The parameters are optimized using optimization algorithms in \textbf{tensorflow}; two main algorithms used in our package are gradient descent and Adam.
Another advantage of using \textbf{tensorflow} is that \textbf{deepspat} employs graphics processing units (GPUs) seamlessly for computation.
The \textbf{deepspat} package is openly available at \url{https://cran.r-project.org/web/packages/deepspat/index.html}, and the code used to reproduce the results in this paper is provided at \url{https://github.com/shaox0a/deepspat_examples}.

\subsection{Functions}

To use the package for fitting and predicting with nonstationary spatial and spatio-temporal processes, users need to specify warping units and model choices.
The warping units are used to construct the layers of the deformation described in \eqref{eq:deformation}. As detailed in Section~\ref{sec:deformations}, the warping units available in the package are the axial warping unit, the radial basis function unit, and the M\"{o}bius transformation unit (also known as a linear fractional transformation, LFT). The \textbf{deepspat} functions that specify the warping units are summarized in Table~\ref{tbl:warping_units}, and their application is demonstrated through a simulation experiment in Section~\ref{sec:simulation} and real-data case studies in Section~\ref{sec:application}. 
Users need to also specify some other model details (including the covariance family on the warped domain, coordinates, and covariates to be inputted into the model) and optimization hyperparameters. Prediction (currently implemented with Gaussian processes only, since this is not often done with models of spatial extremes) can then be performed using the \texttt{predict()} function. These functions are summarized in Table~\ref{tbl:modeling_fns}.

\begin{table}[t!]
	\centering
	\caption{Warping units implemented in \textbf{deepspat}.}
	\label{tbl:warping_units}
	\bgroup
	\def\arraystretch{1}
	\begin{tabular}{ | p{2.5cm} | p{2cm} | p{2.5cm} |  p{9.5cm} |}
		\hline
		  Function & Argument & Default value & Usage \\
		\hline
		\multirow{3}{5cm}{\texttt{AWU()}} & \texttt{r} & 50 & number of basis functions as in \eqref{eq:AWU} \\ & \texttt{dim} & - & dimension of coordinate to apply the AW unit \\ & grad & 200 & steepness parameter $b_1$ in \eqref{eq:AWU_2}  \\
		\hline
		\texttt{RBF\_block()} & \texttt{res} & 1 & resolution of the basis functions used to construct \eqref{eq:RBF}   \\
		\hline
		\texttt{LFT()} & - & - & - \\
		\hline
	\end{tabular}
	\egroup
\end{table}

\begin{table}[ht!]
	\centering
	\caption{Important functions in \textbf{deepspat} and some of their arguments.}
	\label{tbl:modeling_fns}
	\bgroup
	\def\arraystretch{1}
	\begin{tabular}{ | p{5.5cm} | p{2.5cm}  p{9.5cm} |}
        \hline
		  Function & Argument & Usage \\
		\hline\hline
        \multicolumn{3}{|l|}{Functions for Gaussian processes} \\
		\hline
		\multirow{5}{5.5cm}{\texttt{deepspat\_GP(), deepspat\_bivar\_GP()} \\ Usage: Univariate and bivariate Gaussian processes for spatial data} & \texttt{f} & \texttt{R} formula identifying the response variable and the spatial/spatio-temporal coordinates \\ & \texttt{g} & \texttt{R} formula identifying the covariates in the linear trend \\ & \texttt{layers} & warping layers structure   \\ & \texttt{family} & family of the covariance model on the warped domain \\ & \texttt{par\_init, learn\_rates, nsteps} & hyperparameters for optimization \\
		\hline
		\multirow{5}{5.5cm}{\texttt{deepspat\_nn\_GP(), deepspat\_nn\_ST\_GP()} \\ Usage: Nearest neighbor Gaussian processes for spatial/spatio-temporal data} & \multicolumn{2}{ p{12cm}|}{Same arguments as \texttt{deepspat\_GP()}, and} \\ & \texttt{m} & number of nearest neighbors \\ & \texttt{order\_id} & indices of the ordered observations \\ & \texttt{nn\_id} & indices of the nearest neighbors (obtained from the package \textbf{GpGp}) \hspace{6.5cm} \\
        \hline
        \multirow{2}{2cm}{\texttt{predict()}} & \texttt{object} & fitted \textbf{deepspat} model \\ & \texttt{newdata} & new locations at which to predict \\
		\hline\hline
        \multicolumn{3}{|l|}{Functions for extremal processes} \\
		\hline
		\multirow{5}{5cm}{\texttt{deepspat\_MSP()} \\ Usage: max-stable processes for spatial extremes} & \multicolumn{2}{ p{12cm}|}{Same arguments as \texttt{deepspat\_GP()}, and} \\
        & \texttt{nsteps\_pre} & hyperparameters for optimization \\
        & \texttt{method} & estimation method \\
        & \texttt{edm\_emp} & empirical estimate of extremal dependence measures\\
        & \texttt{p} & proportion of subsampling site pairs for pairwise composite likelihood\\
		\hline
        \multirow{5}{5cm}{\texttt{deepspat\_rPP()} \\ Usage: $r$-Pareto processes for spatial extremes} & \multicolumn{2}{ p{12cm}|}{Same arguments as \texttt{deepspat\_MSP()}, and} \\
        & \texttt{risk} & risk functional \\
        & \texttt{weight\_fun} & weight function for gradient score matching \\
        & \texttt{dWeight\_fun} & derivative of the weight function \\ 
        & \texttt{thre} & threshold $u$ \\
        \hline
        \multirow{2}{2cm}{\texttt{summary()}} & \texttt{object} & fitted \textbf{deepspat} model \\ & \texttt{newdata} & new locations at which to construct the extremal dependence pattern \\
        \hline
	\end{tabular}
	\egroup
\end{table}

\subsection{Simulation study}\label{sec:simulation}

We now present a simulation study to demonstrate the use of the \textbf{deepspat} package for modeling nonstationary spatial data. 
We use two datasets: one generated by \textbf{deepspat} and one generated by the package \textbf{cocons} \citep{blasi2024class}. 
To generate a simulated dataset from \textbf{deepspat}, we use the following code,
\begin{verbatim}
deepspat_data <- sim_data(type = "AWU_RBF_2D", ds = 0.01, n = 6000L, sigma2y = 0.01)
\end{verbatim}
where \texttt{ds} denotes the spacing on our $[-0.5,0.5] \times [-0.5,0.5]$ grid, \texttt{n} is the number of observations, and \texttt{sigma2y} denotes the measurement error variance.
From \textbf{cocons}, we use the dataset \texttt{holes} that is included in the package.
Each of the two datasets has $6000$ observations, and we randomly sampled $1500$ observations for training, and left the remaining $4500$ observations for testing. 

We fitted the data using the models introduced in Section~\ref{sec:gaussian_models}, that is, a Gaussian process model, a nearest neighbor Gaussian process model (with the number of neighbors set to 50), and a fixed rank kriging model, implemented through \textbf{deepspat}. We also compared the predictive performance of our package with that from \textbf{cocons}, the model from which was used to generate the dataset \citep[which is based on the framework proposed by][]{paciorek2006spatial}, and with the package \textbf{gstat} \citep{gstat2016}, which is commonly used to fit stationary spatial models.

To fit the models, we first need to specify the structure of the deformations. We set the deformation function to consist of two axial warping units (one for each axis), a radial basis function unit, and a M\"{o}bius transformation unit. In the code below, in the \texttt{AWU} layer, \textsf{r} denotes the number of basis functions, \texttt{grad} represents the parameter $b_1$ in \eqref{eq:AWU_2}, while \texttt{lims} contains the boundaries of the axis identified by \texttt{dim} that is being warped:

\begin{verbatim}
layers_gp <- c(AWU(r = 50L, dim = 1L, grad = 50, lims = c(-0.5, 0.5)),
               AWU(r = 50L, dim = 2L, grad = 50, lims = c(-0.5, 0.5)),
               RBF_block(),
               LFT())
\end{verbatim}

For the fixed rank kriging basis-function model, we also need to specify the structure of the basis functions $\phivec(\cdot)$. Here, we choose 400 bisquare basis functions automatically laid out on a $20 \times 20$ grid by the package, and append them to the warping layers as follows:

\begin{verbatim}
layers <- c(layers_gp,
            bisquares2D(r = 400L))
\end{verbatim}

The next step is to fit the three models by calling the \textbf{deepspat} family of functions in Table~\ref{tbl:modeling_fns}. 
We first convert our data to a data frame with appropriate fields, and split the data into a training set and a test set:
\begin{verbatim}
deepspat_data_all <- data.frame(x = deepspat_data$sobs[,1],
                                y = deepspat_data$sobs[,2],
                                z = deepspat_data$y)
deepspat_data_train <- deepspat_data_all[sample(1:nrow(deepspat_data_all), 1500),]
deepspat_data_test <- setdiff(deepspat_data_all, deepspat_data_train)

\end{verbatim}

The code shown below is used to fit the Gaussian process model; fitting the other two models is done in a similar fashion.

\begin{figure}[]
\begin{center}
\includegraphics[width=0.9\linewidth]{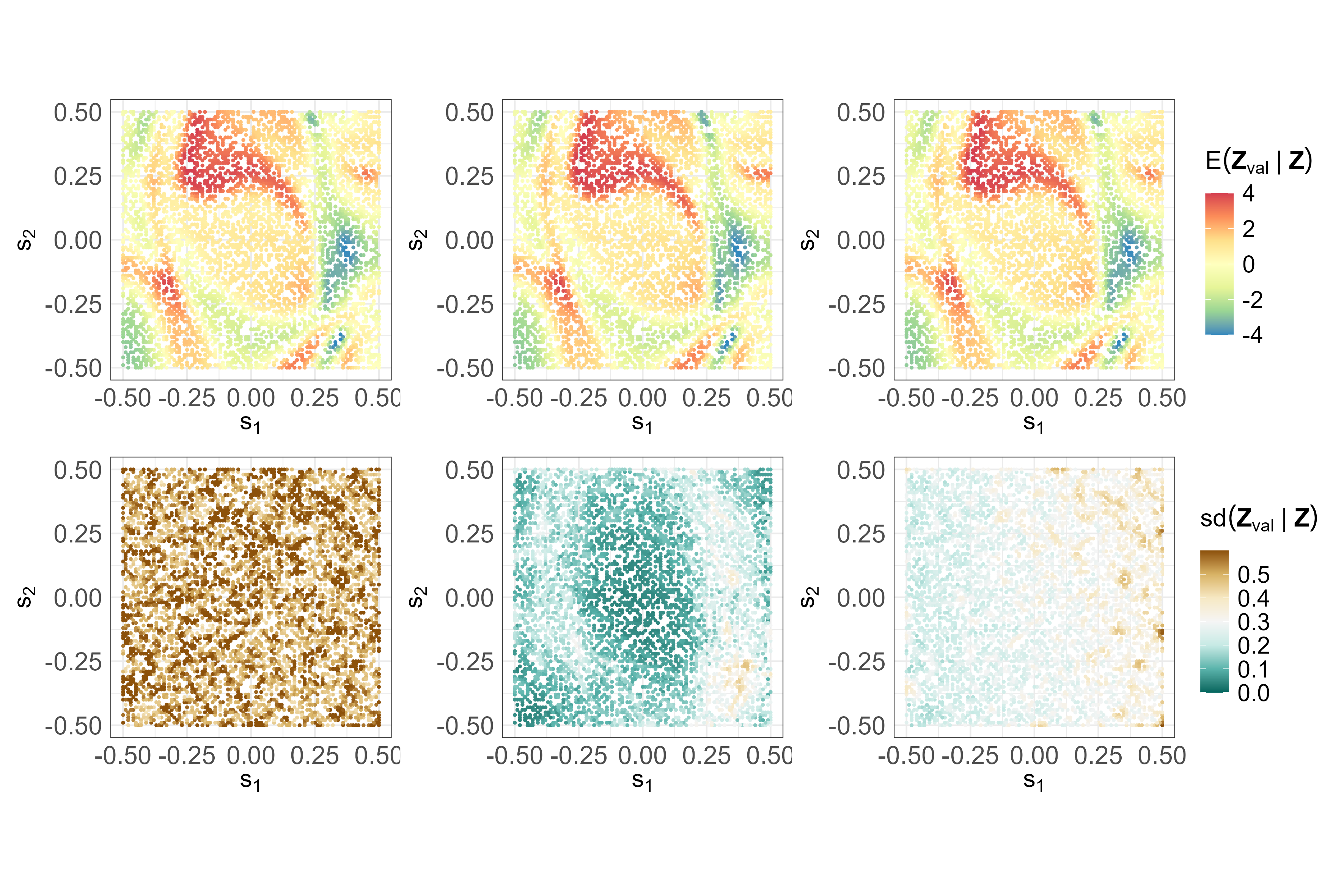}
\includegraphics[width=0.9\linewidth]{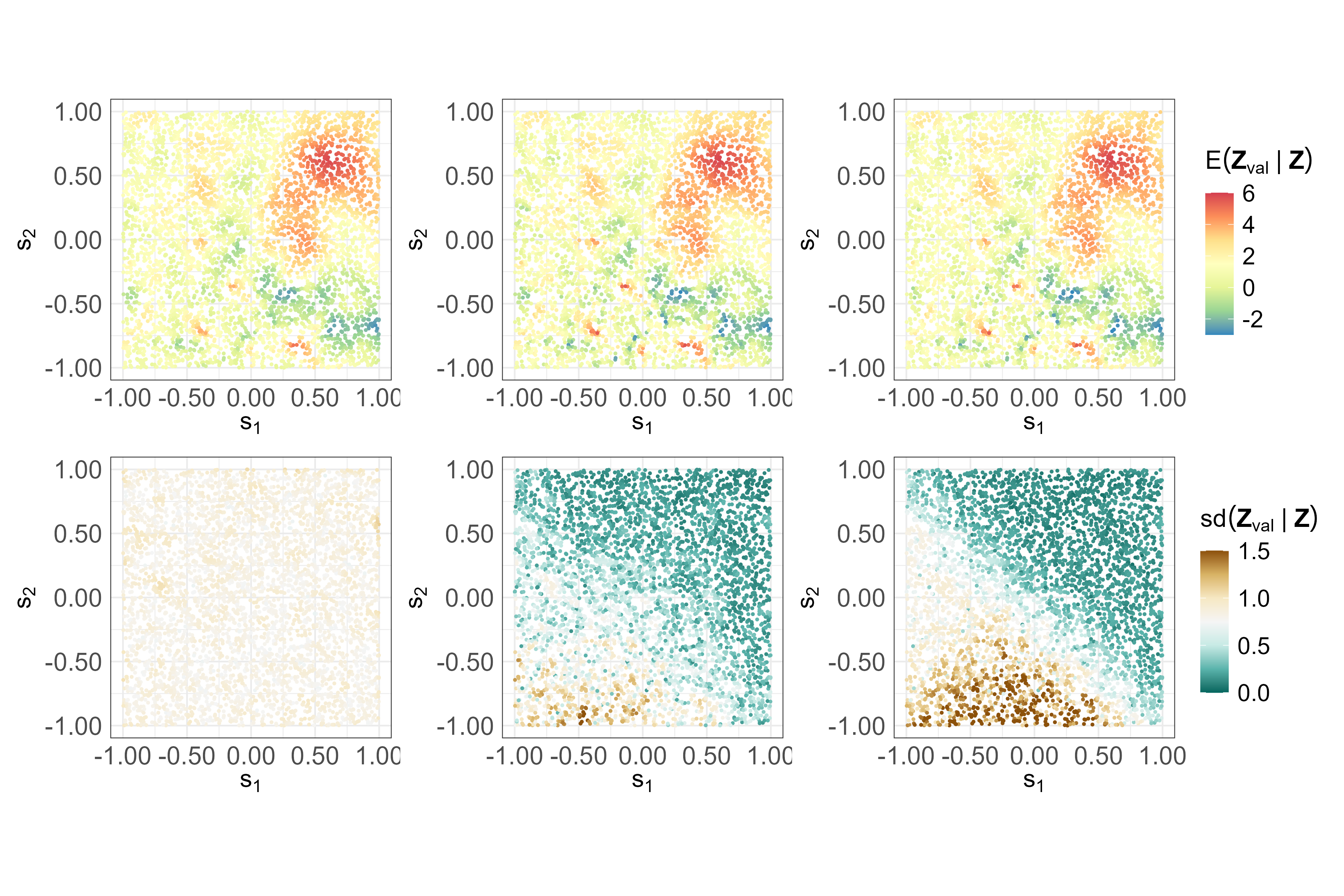}
\end{center}
\caption{Predicted values and prediction standard errors at testing locations for the two simulated datasets. 
First and second rows: Predicted values and prediction standard errors for the simulated data from \textbf{deepspat}. Third and fourth rows: Predicted values and prediction standard errors for the simulated data from \textbf{cocons}. By columns: Prediction from the stationary model fitted using \textbf{gstat}, and the nonstationary Gaussian process models fitted using \textbf{deepspat} and \textbf{cocons}, respectively.}
\label{fig:sim}
\end{figure}

\begin{verbatim}
d_gp <- deepspat_GP(f = z ~ x + y - 1,
                    data = deepspat_data_train,
                    g = ~ 1,
                    layers = layers_gp,
                    family = "exp_nonstat",
                    nsteps = 50L,
                    par_init = initvars(l_top_layer = 0.5),
                    learn_rates = init_learn_rates(eta_mean = 0.02))
\end{verbatim}
In the above code, \texttt{f} is used to specify the spatial coordinates, \texttt{data} specifies the dataframe, \texttt{g} is used to specify the covariates in the mean, \texttt{layers} denotes the warping layers specified above, \texttt{family} denotes the type of nonstationary covariance model used, and \texttt{par\_init} and \texttt{learn\_rates} are optimization hyperparameters which specify initial values and learning rates, with \texttt{l\_top\_layer} and \texttt{eta\_mean} specifically denoting the initial value for the lengthscale of the top layer and the learning rate of the weights in the warping functions. Finally, prediction is performed using the \texttt{predict} function: 
\begin{verbatim}
pred_gp <- predict(d_gp, deepspat_data_test)
\end{verbatim}

We compared the performance of all the models with two scoring rules: the root mean squared prediction error (RMSPE) and the continuous ranked probability score (CRPS). From Table~\ref{tbl:sim1}, we can see that amongst the \textbf{deepspat} models, the vanilla nonstationary Gaussian process model performs best, which is expected since both the nearest neighbor Gaussian process and the fixed rank kriging models introduce approximations. 
For the simulated data from \textbf{deepspat}, all of the \textbf{deepspat} models outperform the nonstationary model implemented in \textbf{cocons}, which itself outperforms the stationary model in \textbf{gstat} in both RMSPE and CRPS. In particular, the CRPS of the nonstationary model from \textbf{cocons} is about $43\%$ higher than the best model from \textbf{deepspat}.
By contrast, for the simulated data from \textbf{cocons}, the nonstationary Gaussian process in \textbf{deepspat} more closely matches the performance of \textbf{cocons}, which was used to generate the data. In particular, the CRPS of the best \textbf{deepspat} model is only about $7\%$ larger than that of \textbf{cocons}.
We also show the predicted values and prediction standard errors from the models in Figure~\ref{fig:sim}, where we see substantial differences in the standard error maps between the stationary and nonstationary models.

\begin{table}[t!]
	\centering
	\caption{Testing results for the simulation study across two datasets. Bolded values indicate best performance for the respective dataset}
	\label{tbl:sim1}
	\bgroup
	\def\arraystretch{1.2}
	\begin{tabular}{ |l|cc|cc| }
		\hline
		\multirow{2}{*}{Model} & \multicolumn{2}{c|}{Dataset from \textbf{deepspat}} & \multicolumn{2}{c|}{Dataset from \textbf{cocons}} \\
		\cline{2-5}
		 & RMSPE & CRPS & RMSPE & CRPS \\
         \hline
        Stationary GP model in \textbf{gstat} & 0.213 & 0.157 & 0.786 & 0.404 \\
        \hline
        Nonstationary GP model in \textbf{cocons} & 0.211 & 0.117 & \textbf{0.729} & \textbf{0.295} \\
        \hline
        Nonstationary NNGP model in \textbf{deepspat} & 0.172 & 0.105 & 0.882 & 0.448 \\
        \hline
        Nonstationary FRK model in \textbf{deepspat} & 0.166 & 0.092  & 0.944 & 0.489 \\
        \hline
	Nonstationary GP model in \textbf{deepspat} & \textbf{0.143} & \textbf{0.082} & 0.757 & 0.316 \\
        \hline
	\end{tabular}
	\egroup
\end{table}

\section{Application to Nepal Temperature Data}\label{sec:application}

\begin{figure}[t!]
\begin{center}
\begin{tabular}{c}
\includegraphics[width=0.5\linewidth]{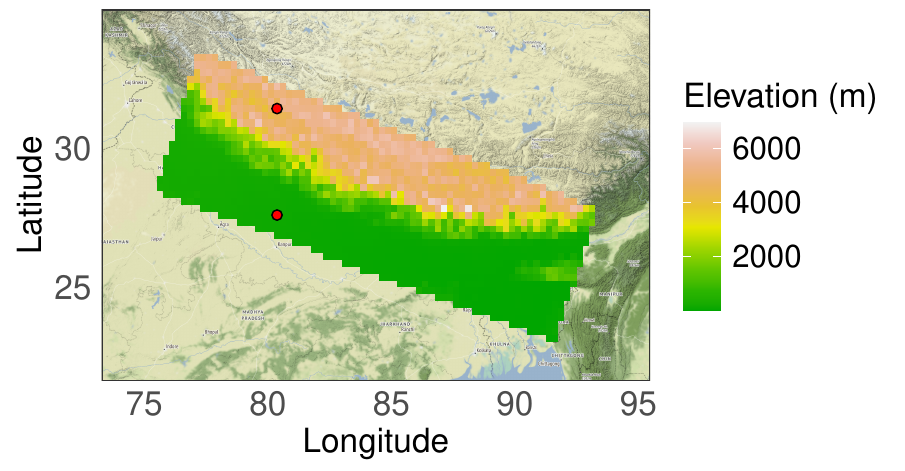}
\end{tabular}
\end{center}
\caption{Elevation of selected spatial domain around Nepal. Two reference sites shown in red are used to analyze inferred spatial dependencies in Sections \ref{sec:app_gaussian} and \ref{sec:app_extremes}.}
\label{pic:app_nepal}
\end{figure}

To illustrate the modeling of nonstationary spatial and spatio-temporal data using \textbf{deepspat}, we consider Nepal temperature data generated by Version 2 of the NASA Global Land Data Assimilation System \citep{rodell2004global}, from 2004 to 2019. We specifically considered $n = 1419$ observation sites on a regular grid within and surrounding Nepal, where there are big changes in elevation; see Figure \ref{pic:app_nepal}. For brevity, from hereon we only show parts of the code needed to reproduce the results; for full reproducible code see the scripts in \url{https://github.com/shaox0a/deepspat_examples}.

\subsection{Annual mean temperature} \label{sec:app_gaussian}

We first investigate the spatio-temporal variation of the mean annual temperature, obtained by taking the mean of the temperature every year at each location from 2004 to 2019. This mean annual temperature dataset has, in total, 22,704 data points.
We randomly sample 50\% of these data points and use them for fitting the model; we leave the remaining 50\% for testing. 

To these data, we fit a spatio-temporal model with a spatial warping function consisting of two axial warping units (one for each spatial dimension), a radial basis unit, and a M\"{o}bius transformation unit. For the time dimension, we use a single axial warping unit. 
We use an exponential covariance function approximated by a nearest-neighbor Gaussian process on the warped domain.
To set up the model, the first step is to specify the spatial and temporal warping layers:
\begin{verbatim}
layers_spat <- c(AWU(r = 100L, dim = 1L, grad = 200),
                 AWU(r = 100L, dim = 2L, grad = 200),
                 RBF_block(res = 1L), LFT())
layers_temp <- c(AWU(r = 20L, dim = 1L, grad = 20))
\end{verbatim}
As we use the nearest neighbor GP approximation, we also need to specify the ordering and neighbor structure. Below, we choose a random ordering, and set the number of neighbors to 50:
\begin{verbatim}
locs_train <- as.matrix(obsdata[c("s1", "s2", "year")])
order_id <- sample(1:nrow(locs_train))
nn_id <- find_ordered_nn(locs_train[order_id,], m = 50)
\end{verbatim}
Then, we fit the nonstationary spatio-temporal model using \texttt{deepspat\_nn\_ST\_GP()}. The input into the warping includes the spatial coordinates (longitude and latitude), and the year. For covariates, we consider both an intercept-only model (for which the code is shown below) and an intercept plus elevation model: 
\begin{verbatim}
d_st_gp <- deepspat_nn_ST_GP(f = Y_mean ~ s1 + s2 + year - 1, 
                             g = ~ 1,
                             data = obsdata, 
                             family = "exp_nonstat_sep",
                             layers_spat = layers_spat, 
                             layers_temp = layers_temp,
                             m = 50L, 
                             nsteps = 50L,
                             order_id = order_id, 
                             nn_id = nn_id,
                             par_init = initvars(l_top_layer = 0.1),
                             learn_rates = init_learn_rates(eta_mean = 0.003,
                                                            LFTpars = 0.001)
                             )
\end{verbatim}
Finally, we make predictions using the \texttt{predict()} function. Since we are employing a nearest neighbor model, here we also need to specify the neighbors of each prediction location (which can include the observation locations). We get the nearest neighbors using the function \texttt{get.knnx} from the package \textbf{FNN}. Here we choose the 50 nearest neighbors for prediction.
\begin{verbatim}
locs_new <- as.matrix(alldata[c("s1", "s2", "year")])
nn_id_pred <- FNN::get.knnx(data = locs_train, query = locs_new, k = 50)$nn.index
pred <- predict(d_st_gp, alldata, nn_id_pred)
\end{verbatim}

\begin{figure}[t!]
\begin{center}
\begin{tabular}{c}
\includegraphics[width=0.9\linewidth]{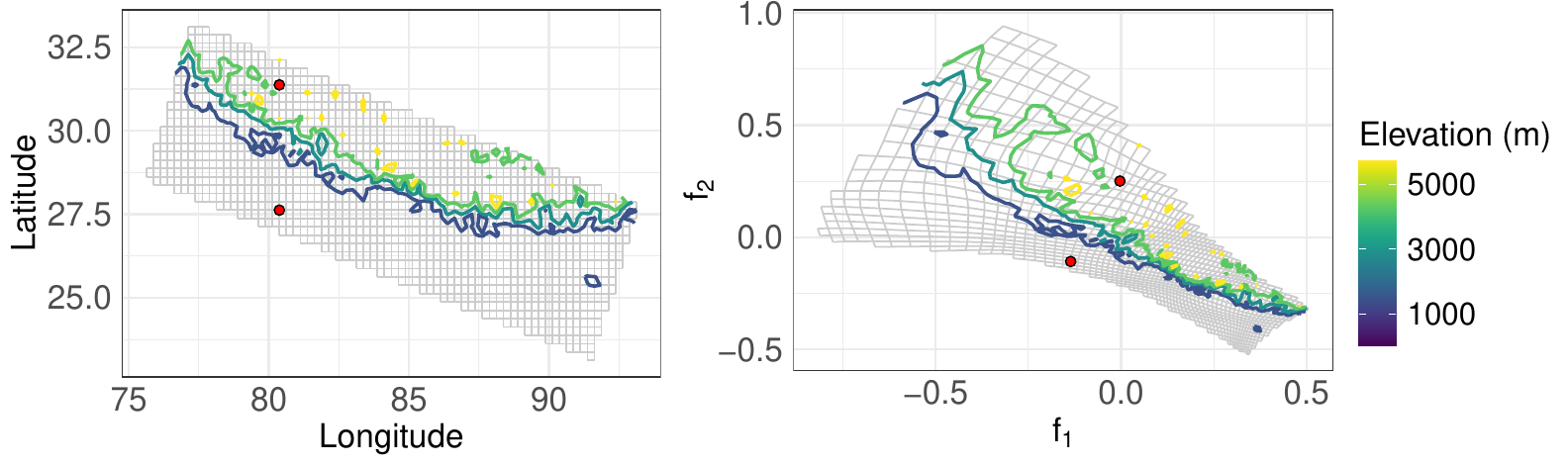}
\end{tabular}
\end{center}
\caption{Original (left) and warped (right) spaces obtained by modeling annual mean temperature, using a nonstationary spatio-temporal model and nearest neighbor Gaussian processes.}
\label{fig:gaussian_estimated_space}
\end{figure}

\begin{figure}[t!]
\begin{center}
\begin{tabular}{c}
\includegraphics[width=0.95\linewidth]{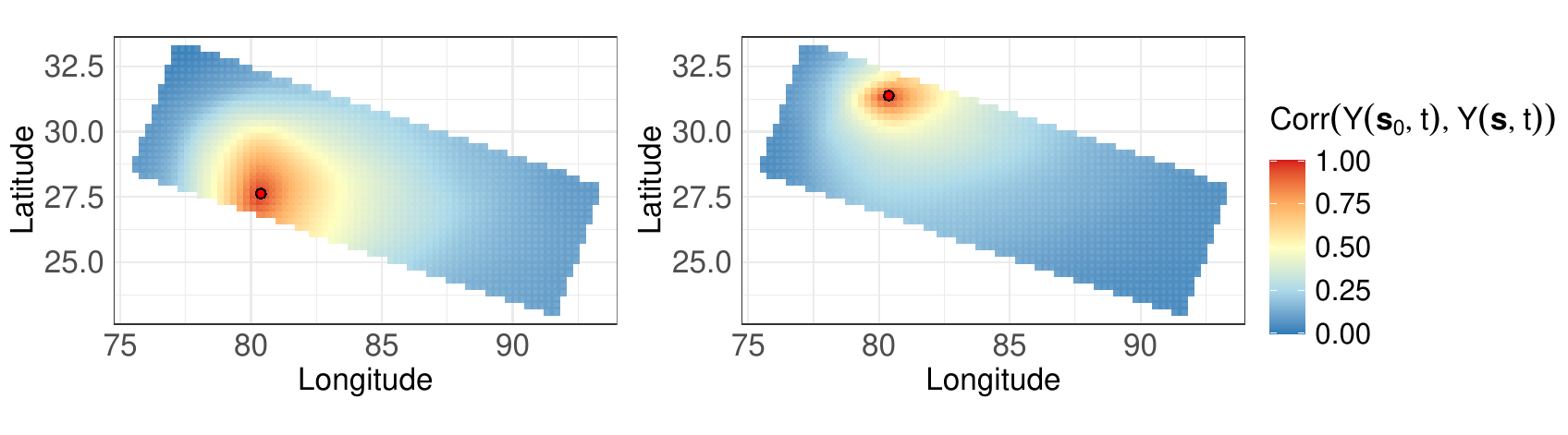}
\end{tabular}
\end{center}
\caption{Spatial correlation between the reference site $\svec_0$ (labeled in red) and other locations $\svec \in \mathcal{S}$, denoted by $\operatorname{Corr}(Y(\svec_0; t), Y(\svec; t))$, for $\svec_0$ in plain area (left panel) and $\svec_0$ in the mountain range (right panel) at any time point $t$ in the time horizon of the study.}
\label{fig:gaussian_estimated_corr}
\end{figure}

The original space $\mathcal{S}$ and the inferred warped space $\mathcal{W}$ are shown in Figure~\ref{fig:gaussian_estimated_space}. 
We see that space is contracted toward the east and expanded toward the west.
The estimated spatial covariance structure can be visualized using a correlation heat map, which shows the correlation of the process at a reference location with the process at other locations. Figure~\ref{fig:gaussian_estimated_corr} shows the correlation heat map from the nonstationary model that includes elevation as a covariate for two different reference locations: one in the mountain range and one at lower elevation. It can be seen that the correlation structure is different at these two locations (showing nonstationarity), and that the correlation decreases faster for the high-elevation reference location.
Finally, the predictive performance of \textbf{deepspat} models is summarized in Table~\ref{tbl:pred_performance}. The nonstationary model with elevation as a covariate has substantially lower RMSPE and CRPS than the other models, indicating a better fit.

\begin{table}[t!]
	\centering
	\caption{Predictive performance of spatio-temporal models in \textbf{deepspat}, assessed using predictive metrics on left-out testing data. Bolded values indicate best performance.}
	\label{tbl:pred_performance}
	\bgroup
	\def\arraystretch{1}
	\begin{tabular}{ |c|cc| }
		\hline
		Model & RMSPE & CRPS \\
		\hline
		Stationary, intercept-only & 0.1184 & 0.1019   \\
		\hline
		Stationary, with elevation & 0.0935 & 0.0972   \\
		\hline
		Nonstationary, intercept-only & 0.0942 & 0.0470  \\
		\hline
		Nonstationary, with elevation & \textbf{0.0739} & \textbf{0.0423}   \\
		\hline
	\end{tabular}
	\egroup
\end{table}

\subsection{Monthly maximum temperature} \label{sec:app_extremes}


We now present the flexibility of our models in capturing nonstationary extremal dependence structure using max-stable processes, where extremes are defined as monthly maxima. For modeling of the extremal dependence, we here leverage a two-stage procedure: we first model the marginal behavior of the data before transforming it onto common unit Fréchet, and then model the extremal dependence structure using these standardized data. We compute monthly maxima at each spatial location and obtain 192 fields, which we treat as independent. 

Similar to the modeling of annual mean temperature, we fit a spatial model whose warping function consists of two axial warping units (one for each spatial dimension), a single resolution radial basis unit with rank 1, and a M\"{o}bius transformation unit, combined with the power variogram. We first specify the spatial layers by:

\begin{verbatim}
layers_spat2 <- c(AWU(r = 50L, dim = 1L, grad = 200, lims = c(-0.5, 0.5)),
                AWU(r = 50L, dim = 2L, grad = 200, lims = c(-0.5, 0.5)),
                RBF_block(1L), 
                LFT())
\end{verbatim}

We then fit the nonstationary spatial models for extremes using the function \texttt{deepspat\_MSP()} using the randomized pairwise likelihood approach (denoted by \texttt{MRPL}):

\begin{verbatim}
fit_msp_nonstat <- deepspat_MSP(
                   f = as.formula(paste(paste(paste0("z", 1:(ncol(obs_all) - 2)),
                   collapse = "+"), "~ s1 + s2 - 1")),
                   data = obs_all,
                   layers = layers_spat2,
                   method = "MRPL",
                   family = "power_nonstat",
                   dtype = "float64",
                   nsteps = 50L,
                   nsteps_pre = 50L,
                   par_init = initvars(),
                   learn_rates = init_learn_rates(eta_mean = 0.01,
                                                  vario = 0.01),
                   edm_emp = obs_edm_emp,
                   p = 0.01)
\end{verbatim}
The extremal dependence pattern at unobserved locations can then be obtained by calling the \texttt{summary()} function: 
\begin{verbatim}
summ <- summary(fit_msp_nonstat, df_loc)
\end{verbatim}

\begin{figure}[t!]
\begin{center}
\begin{tabular}{c}
\includegraphics[width=0.9\linewidth]{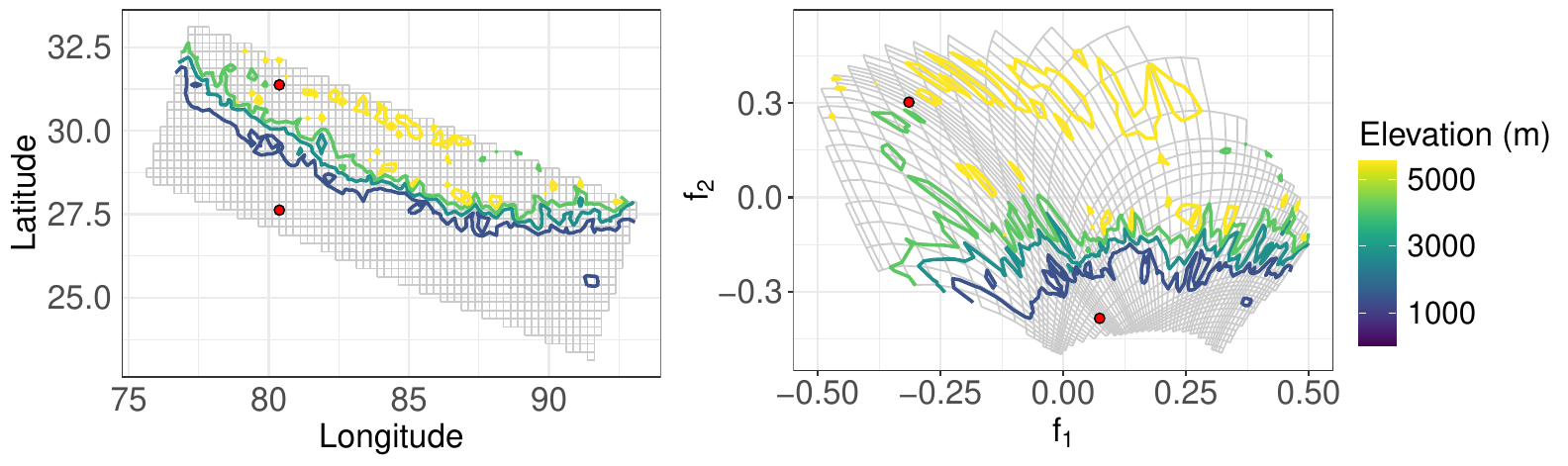} \\
\end{tabular}
\end{center}
\caption{Original (left) and warped (right) spaces obtained by modeling monthly maxima, using a nonstationary max-stable process.}
\label{pic:nepal_ext_space}
\end{figure}

\begin{figure}[hbt!]
\begin{center}
\begin{tabular}{c}
\includegraphics[width=0.8\linewidth]{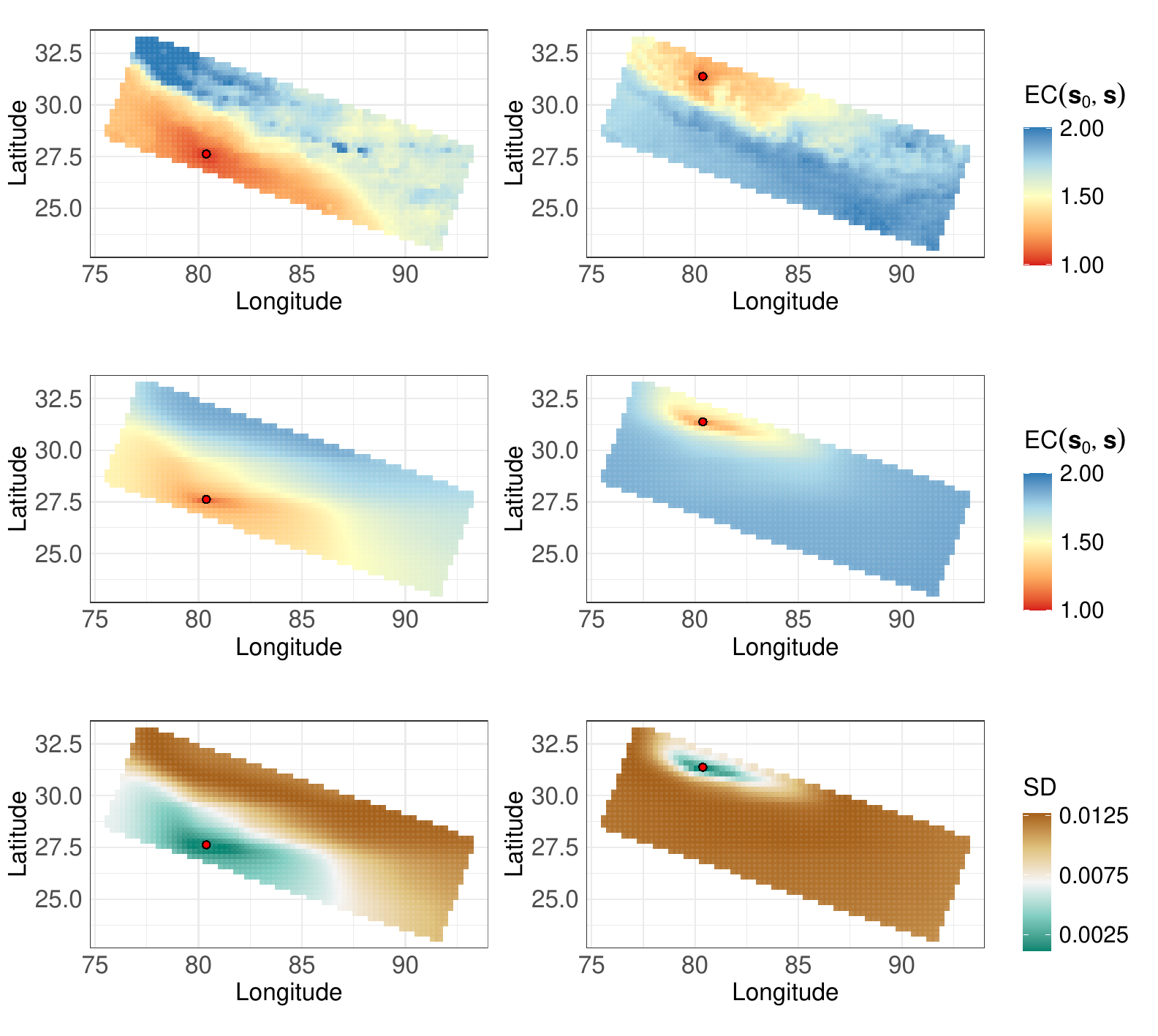} \\
\end{tabular}
\end{center}
\caption{Empirical estimates (top row) and model-based estimates (middle row) of pairwise extremal coefficients (EC) between the reference site $\svec_0$ (labeled in red) and other sites $\svec \in \mathcal{S}$, denoted as $\operatorname{EC}(\svec_0, \svec)$, and the standard deviation of the corresponding model-based EC estimates (bottom row), approximated using the Delta method (with the estimated warping function treated as fixed), for $\svec_0$ in the low-elevation area (left column) and $\svec_0$ in the mountain range (right column).}
\label{pic:nepal_ext_ec}
\end{figure}

\begin{figure}[t!]
\begin{center}
\begin{tabular}{c}
\includegraphics[width=0.5\linewidth]{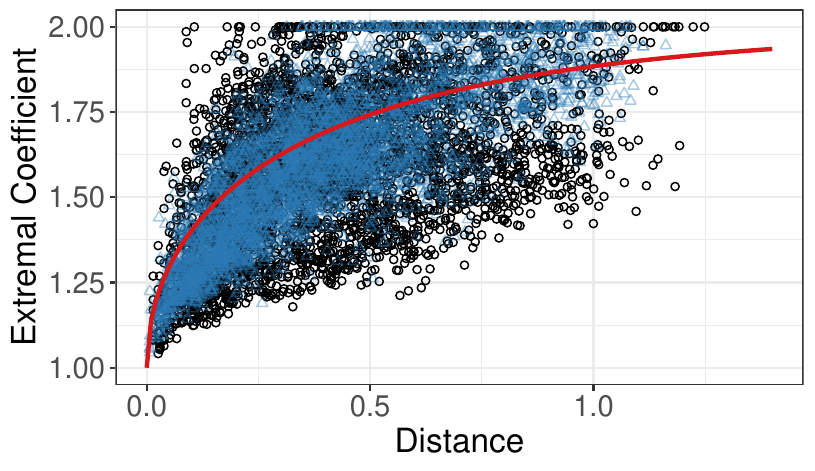}
\end{tabular}
\end{center}
\caption{Pairwise extremal coefficients against distances plotted between site pairs in the original (black) and warped spaces (blue), along with the model fitted extremal coefficient curve (red).}
\label{pic:nepal_ext_cloud}
\end{figure}

To visualize the fitted models, we show the estimated warping in Figure~\ref{pic:nepal_ext_space}, and the estimated pairwise extremal dependence structure relative to the two reference sites, along with standard deviations of the estimates, in Figure~\ref{pic:nepal_ext_ec}. 
In the warped space, a notable contraction occurs in low-elevation regions, suggesting stronger extremal dependence among observations in those areas. This nonstationary pattern is further supported by both empirical and model-based estimates of pairwise extremal coefficients (see Figure~\ref{pic:nepal_ext_ec}). 
We quantify uncertainty in the warped space using a multivariate Delta-method approximation around the estimated dependence parameters while holding the estimated warping fixed. These approximations are obtained outside the package and can be viewed in our supplied reproducible code. 
Estimated extremal coefficients have smaller uncertainty for nearby (strongly dependent) site pairs and larger uncertainty for distant (weakly dependent) pairs.
In Figure~\ref{pic:nepal_ext_cloud}, we plot the extremal coefficients against distance between location pairs. The close alignment of the points corresponding to the extremal coefficients in the warped space (blue points) with the model-based estimates (red line) provides additional evidence of the model’s ability to capture the complex, nonstationary dependencies in spatial extreme events.

\section{Conclusion}\label{sec:conclusion}

In this paper, we introduce the \textbf{deepspat} package for modeling nonstationary spatial and spatio-temporal data. In the package, nonstationary models are constructed via deep multi-layered warping functions that map the original spatial or spatio-temporal coordinates to new coordinates. These new coordinates are then used for fitting familiar stationary models to Gaussian data (with Gaussian processes or basis function models), or to extremes data (with max-stable processes or $r$-Pareto processes). Fitting of the models is facilitated through gradient-based optimization via \textbf{tensorflow}.
Through a simulation study and a case study of temperature in Nepal, we demonstrate that the models in the package are easy to implement, interpretable, and able to fit complex data well.

There are a number of avenues for package development. 
One of these relates to the development of more general warping functions. Currently, the warping units allowed in the package are those given in Table~\ref{tbl:warping_units}, and are mainly suitable for a two-dimensional spatial domain. There are alternatives for higher-dimensional Euclidean spaces \citep{nag2025modeling} and more complicated manifolds such as spheres \citep{ng2022spherical}, but considerable work will be required to implement these in a robust manner.
Another avenue is to include several other spatial and spatio-temporal models that can be used on the warped domain in the package. We currently limit ourselves to the most common covariance-based models in the Matérn class for Gaussian processes, and the most classical asymptotic extreme-value models.
Finally, it would be useful to extend the package to a generalized linear mixed model setting; doing so will enable it to be used with a wider class of spatial and spatio-temporal non-Gaussian data.

\section*{Acknowledgements}

This material is based upon work supported by the Air Force Office of Scientific Research under award number FA2386-23-1-4100 (A.Z.-M.).

\bibliographystyle{apalike}
\bibliography{biblio}


\end{document}